\documentclass[11pt]{article}
\usepackage{amsmath,amstext,amsbsy,amssymb}
\usepackage{bm}
\usepackage{setspace}
\newcommand{\dso}{\Delta_{\scriptscriptstyle{ SO}}}
\newcommand{\vF}{v_{\scriptscriptstyle{ F}}}
\newcommand{\diag}{{\rm diag}}
\newcommand{\Tr}{{\rm Tr}}

\newcommand{\ssAH}{{\scriptscriptstyle{ AH}}}

\newcommand{\ssR}{{\scriptscriptstyle{ R}}}
\newcommand{\ssH}{{\scriptscriptstyle{ H}}}

\newcommand{\ssK}{{\scriptscriptstyle{ K}}}
\newcommand{\ssKprime}{{\scriptscriptstyle{ K'}}}

\newcommand{\ssSH}{{\scriptscriptstyle{ SH}}}
\newcommand{\ssSO}{{\scriptscriptstyle{ SO}}}

\long\def\symbolfootnote[#1]#2{\begingroup%
\def\thefootnote{\fnsymbol{footnote}}\footnote[#1]{#2}\endgroup}
\begin{document}


\begin{center}

{\Large \bf   

\vspace{2cm}

Relation Between the Spin Hall Conductivity  and the Spin Chern Number for Dirac-like Systems 
}

\vspace{2cm}

\"{O}mer F. Dayi and Elif Yunt

\vspace{5mm}

{\em {\it Physics Engineering Department, Faculty of Science and
Letters, Istanbul Technical University,\\
TR-34469, Maslak--Istanbul, Turkey}}\footnote{{\it E-mail addresses:} dayi@itu.edu.tr , yunt@itu.edu.tr }

\end{center}

\vspace{2cm}

\noindent
A semiclassical formulation of the spin Hall effect for physical systems satisfying Dirac-like  equation  is introduced. We demonstrate that the main contribution to the spin Hall conductivity is given by the spin Chern number whether the spin is conserved  or not at the quantum level. We illustrated the formulation within the Kane-Mele model of graphene in the absence and in the presence of the Rashba spin-orbit coupling term.

\vspace{2.5cm}

\section{Introduction} 
In response to the electric field in ferromagnets a spontaneous Hall current can be generated.  A semiclassical formulation of  this anomalous Hall effect was  given in \cite{haldane} within the Fermi liquid theory. There the anomalous Hall conductivity was calculated considering the equations of motion in the presence of the Berry gauge fields derived from the Bloch wave function. When this system is subjected to an external magnetic field definition of the particle density and the electric current should be done appropriately. Nevertheless the computed value of the anomalous Hall conductivity remains unaltered \cite{xsn}. 
Hall currents without a magnetic field can  be generated
also in fermionic systems described by Dirac-like Hamiltonians \cite{hal88}. Taking into account the spin of electrons these systems yield Hall currents due to the spin transport which is known as the spin Hall effect \cite{km} or Chern insulator.  
We would like to present a semiclassical formulation of the spin Hall conductivity  using  a differential form formalism for fermions which are described by Dirac-like Hamiltonians.

In semiclassical  kinetic theory the spin degrees of freedom can be considered by treating them as  dynamical variables. However to calculate the spin Hall conductivities it would be more appropriate to keep the Hamiltonian and the related Berry gauge fields as  matrices in ``spin indices".
In this respect a differential form formalism was presented in \cite{de}. Dynamical variables in this semiclasical formalism are the usual space coordinates and momentum  but the symplectic form is matrix valued. We will show that this formalism is suitable to calculate the spin Hall conductivity
for Dirac-like systems. We deal with electrons, so that without loss of generality we consider the third component of spin denoted by $S_z,$ whose explicit form depends on the details of the underlying Dirac-like Hamiltonian. When the third component of spin is conserved at the quantum
level, constructing spin current is not intriguing. However spin Hall effect can persist even if the third component of spin is not conserved. In the latter case semiclassical definition of the spin Hall conductivity is not very clear. Within the 
Kane-Mele model of graphene ($2+1$ dimensional topological insulator) \cite{km} it was argued that one cannot anymore use the Berry curvature to obtain the main contribution to the spin Hall effect when the spin nonconserving Rashba term is present \cite{culetal}. We will show that even for the systems where the spin is not a good quantum number it is always possible to establish the leading contribution to the spin Hall effect in terms of the Berry field strength derived in the appropriate basis. Moreover, we will  demonstrate that it is always given in terms of the spin Chern number which is defined to be one half the difference of the Chern numbers of spin-up and spin-down sectors \cite{prodan}. A similar claim was made in \cite{ezawa} by employing the Green function within the Kubo formalism. 

The formulation  will be illustrated 
within the Kane-Mele model of graphene:
When only the intrinsic spin-orbit coupling is present  the third component of the spin is a good quantum number and the spin Hall conductivity   
can be acquired straightforwardly in terms of the Berry curvature \cite{dayiyunt}.   When  the Rashba term is switched on the third component of spin ceases to be   conserved. Nevertheless, we will show that by choosing the correct basis one can still establish the leading contribution to the spin Hall conductivity by the
Berry curvature. It is given by the spin Chern number calculated in \cite{yetal}. 

The starting point of the method is the matrix valued symplectic form \cite{de,ofd}. We will show in Appendix A that it can be obtained 
in terms of the wave packets formed by the positive energy solutions of Dirac-like equations 
adapting  the formalism of \cite{sundaramniu, Culcer}.

The formalism of deriving  the velocities of phase space variables in terms of the phase space variables themselves will be presented in Section \ref{SemiclassicalFormalism}. It  leads to the anomalous  Hall effect straightforwardly as we will discuss  briefly in Section \ref{anoHal}. Definition of the spin current  is presented in Section \ref{SpinCurrent}. It is shown that if one adopts the correct definition of the spin current in two space dimensions  the essential part of the spin Hall conductivity is always given by the spin Chern number. We will illustrate the method   by applying it to the Kane-Mele model first in the absence and then in the presence of Rashba coupling in Section \ref{KM}. The last section is devoted to discussions of the results obtained.

\renewcommand{\theequation}{\thesection.\arabic{equation}}
\section{Semiclassical Formalism}
\label{SemiclassicalFormalism}

We deal with electrons  which effectively satisfy  Dirac-like equations. In Appendix A we presented the semiclassical theory  established in terms of the wave packet composed of positive energy solutions. It  yields a semiclassical description of the system whose dynamics is governed by gauge fields  which are matrices labeled by ``spin indices".  It is so called because the basis of the wave packets are solutions of a Dirac-like Hamiltonian. Obviously range of this index  depends on the spacetime dimension as well as on the intrinsic properties of the system considered. 

In d+1 dimensions, we deal with the following matrix valued one-form,
\begin{equation}
\eta_{\ssH}= p_adx^a+[ea_a^{ext}(x,t)-a_a(x,p)]dx^a-A^a(x,p)dp_a-H(x,p)dt. \nonumber
\end{equation}
 $(x^a,p_a);\ a =1,2,...,d,$ denote the classical phase space variables and  $e>0,$ is the electron charge. $a_a(x,p,t)$ and $ A^a(x,p,t)$ 
are the matrix-valued Berry gauge potentials. $ H(x,p)= H_0(\bm p)-ea_0^{ext} (\bm x)$ comprises of $ H_0$ which is the diagonalized Dirac-like free Hamiltonian projected on  positive energies and the  electromagnetic scalar field  $a_0^{ext}.$ We suppress the unit matrices. The related  symplectic two-form is defined by
\begin{eqnarray}
\label{wgeneral}
w_{\ssH} &=& d\eta_{\ssH}-i\eta_{\ssH} \wedge \eta_{\ssH} \nonumber \\
               &=& dp_a \wedge dx^a -F -G-M - \left(e\frac{\partial a_a^{ext}}{\partial t}+\frac{\partial H}{\partial x^a}-i[H, a_a]\right) dx^a \wedge dt \\
			&&				- \left(\frac{\partial H}{\partial p_a}-i[H, A^a]\right) dp_a \wedge dt. \nonumber
\end{eqnarray}
 For $a_a=0,$ this coincides with the matrix-valued two form considered in \cite{de}.
$F=\frac{1}{2}F_{ab}dx^a \wedge dx^b,\ G=\frac{1}{2}G_{ab}dp^a \wedge dp^b,$ and $M=\frac{1}{2}M_{ab}dp^a \wedge dx^b$ 
are the two-forms with the following components,
\begin{eqnarray}
F_{ab} &=& \frac{\partial a_b}{\partial x^a}- \frac{\partial a_a}{\partial x^b} -i[a_a,a_b]- e\left(\frac{\partial a^{ext}_b}{\partial x^a}- \frac{\partial a^{ext}_a}{\partial x^b}\right), \nonumber \\
{M^a}_b &=& \frac{\partial a_b}{\partial p_a}- \frac{\partial A^a}{\partial x^b} -i[A^a,a_b], \nonumber  \\
G^{ab} &=&\frac{\partial A^b}{\partial p_a}- \frac{\partial A^a}{\partial p_b} -i[A^a,A^b]. \nonumber 
\end{eqnarray}
In order to obtain the equations of motion, we introduce the  matrix valued vector field
\begin{equation}
\label{vf}
\tilde v= \frac{\partial}{\partial t}+\dot{\tilde x}^a\frac{\partial}{\partial x^a}+\dot{\tilde p}_a\frac{\partial}{\partial p_a}.
\end{equation}
Here, $(\dot{\tilde x}^a, \dot{\tilde p}_a)$ are the matrix-valued time evolutions of the phase space variables $(x^a, p_a).$
This is analogous to the situation in the canonical formulation of the Dirac particle where the velocities are matrices although the phase space variables are ordinary vectors. 
The equations of motion are derived by demanding that the interior product of $w_{\ssH},$ (\ref{wgeneral}), with the matrix-valued vector field
 $\tilde v,$  (\ref {vf}), vanish:
\begin{equation}  
i_{\tilde v}w_{\ssH}=0 . \nonumber
\end{equation}
The resulting equations are
\begin{eqnarray}
\dot{\tilde p}_a &=&-\dot{\tilde x}^c F_{ca}+e_a+{M^c}_a\dot{\tilde p}_c ,\nonumber \\
\dot{\tilde x}^a &= &- G^{ca}\dot{\tilde p}_c-f^a+\dot{\tilde x}^c{M^a}_c,\nonumber 
\end{eqnarray}
where in terms of the external electric field ${\cal E}_a= \partial a_0^{ext}/ \partial x^a- \partial a^{ext}_a / \partial t,$ we defined
\begin{eqnarray*}
e_a&\equiv&e{\cal E}_a -i[H_0, a_a] ,\\
f^a&\equiv& -\frac{\partial H_0}{\partial p_a}-i[H_0, A^a].
\end{eqnarray*}
The Lie derivative of the volume form  $\Omega_{d+1}=\frac{(-1)^{[\frac{d}{2}]}}{d!}w^d_{\ssH} \wedge dt$ can be used to attain the matrix-valued velocities $(\dot{\tilde x}^a,\dot{\tilde p}_a)$ in terms of the phase space variables $(x^a,p_a).$ We will illustrate it for $d=2,$
since  basically we are interested in  $2+1$ spacetime dimensional Dirac-like systems.
In $2+1$ dimensions, where the extended phase space is 5 dimensional, the volume form reads
\begin{equation}
\label{vf21}
\Omega_{2+1}=-\frac{1}{2}w_{\ssH} \wedge w_{\ssH} \wedge dt .
\end{equation}
We express it  through  the canonical volume element of the phase space $dV$, as 
\begin{equation}
\label{volf}
\Omega_{2+1}=\tilde{w}_{1/2}dV\wedge dt ,
\end{equation}
where $\tilde{w}_{1/2}$  is  the  Pfaffian of the following $4\times 4$ matrix,
$$
\begin{pmatrix}
-{F}_{ij} &-{\delta^i}_j+{M^i}_j\\
{\delta^i}_j-{M^i}_j & G^{ij}
\end{pmatrix} .
$$
We do not treat the spin indices on the same footing with the phase space indices $(x_i, p_i);\ i=1,2.$
Thus the Pfaffian $\tilde{w}_{1/2}$
is a matrix in spin indices.
The Lie derivative associated with the vector field  (\ref {vf}) of the volume form  (\ref {volf}) can  be expressed formally as 
\begin{equation}
\label{lieomega1}
L_v\Omega_{2+1}=(i_vd+di_v)(\tilde{w}_{1/2}dV\wedge dt)=\left(\frac{\partial}{\partial t}\tilde{w}_{1/2}+\frac{\partial}{\partial x^i}(\dot{\tilde x}^i\tilde{w}_{1/2})+\frac{\partial}{\partial p_i}(\tilde{w}_{1/2}\dot{\tilde p}_i)\right)dV\wedge dt.
\end{equation}
Actually, to obtain  it explicitly one should employ the  definition (\ref{vf21}) yielding
\begin{equation}
\label{lieomega2}
L_v\Omega_{2+1}=-\frac{1}{2}dw^2_{\ssH}. \nonumber
\end{equation}
Comparing the  exterior derivative of
\begin{eqnarray}
 w_{\ssH} \wedge w_{\ssH} 
                   &=& dp_i \wedge dx^i \wedge dp_j \wedge dx^j- 2M \wedge dp_i \wedge dx^i+ 2e_i dx^i \wedge dt \wedge dp_j \wedge dx^j \nonumber \\
                   &-&\frac{1}{2}\{F,f_i\}\wedge dp^i \wedge dt -\{M,e_i\} \wedge dx^i \wedge dt \nonumber \\
                   &+& 2f^i dp_i \wedge dt \wedge dp_j \wedge dx^j +F \wedge G +G \wedge F - \{M,f^i\} \wedge dp_i \wedge dt \nonumber \\
                   &-& \frac {1}{2}\{G,e_i\} \wedge dx^i \wedge dt, \nonumber 
\end{eqnarray}
with the formal expression (\ref{lieomega1}) one obtains the solutions
\begin{eqnarray}
\tilde{w}_{1/2} &=& 1+{M^i}_{i}-\frac{1}{4}\{F_{ij},G^{ij}\} ,\label{eom1}\\
\dot{\tilde x}^i\tilde{w}_{1/2} &=& -f^i-\{{M^i}_j,f^j\}+\{{M^j}_j,f^i\}+\frac{1}{2}\{G^{ij},e_j\} ,\label{eom2} \\ 
\label{eom3}
\tilde{w}_{1/2}\dot{\tilde p}_i &=& e_i+\{{M^j}_i,e_j\}-\{{M^j}_j,e_i\}-\frac{1}{2}\{F_{ji},f^j\} .
\end{eqnarray}
These solutions are useful even for  Schr\"{o}dinger type Hamiltonian systems where the origin of the Berry gauge fields will be different.
Indeed, to illustrate the power of the differential form method in general
 we would like to deal briefly with the anomalous Hall effect in two dimensions.

\setcounter{equation}{0}

\section{Anomalous Hall Effect}
\label{anoHal}

The intrinsic anomalous Hall effect in ferromagnetic materials arises from the Berry curvature in the crystal momentum space  of Bloch electrons
either in the absence  or in the presence of an external magnetic field \cite{haldane,xsn}. In the latter case one should define the electric current 
by taking  corrections to the path integral measure  into account. 
The anomalous Hall conductivity can be derived within  the formalism of Section \ref{SemiclassicalFormalism}. Obviously, in this case the Berry gauge fields are derived from the occupied Bloch states.   Consider the  electrons which are constrained to move in the $xy$-plane in the presence of the constant magnetic field in the $z$-direction $ F_{xy}=B,$ and the Berry curvature $G^{xy}.$ The equations of motion (\ref{eom1})-(\ref{eom3}) become
\begin{eqnarray}
\sqrt{w} &=& 1+BG^{xy},\label{eomah1}\\
\sqrt{w}\dot{x}^i &=& \frac{\partial H}{\partial p_i}+e\epsilon^{ij}{\cal E}_jG^{xy} , \label{eomah2} \\ 
\sqrt{w}\dot{p}_i &=& e {\cal E}_i +\epsilon_{ij}\frac{\partial H}{\partial p_j}B.
\end{eqnarray}
 (\ref{eomah1}) is the correction to the path integral measure. Here $\epsilon_{ij}=g_{ik}g_{jn}\epsilon^{kn}$, where $g_{ik}$ is the Euclidean metric. Hence, the correct definition of electric current is
 $$ j^i=e\int \frac{d^2p}{(2\pi\hbar)^2}\sqrt{w}\dot{x}^i f (x,p,t),$$ 
where $f (x,p,t)$ is the ground state distribution (occupation) function.
Plugging (\ref{eomah2}) into this definition one obtains  the total electric current as
\begin{equation}
j^i=e\int \frac{d^2p} {(2\pi\hbar)^2} \left( \frac{\partial H}{\partial p_i}+e\epsilon^{ij}{\cal E}_jG^{xy} \right)f (x,p,t). \nonumber
\end{equation}
The term proportional to the electric field yields the anomalous Hall current
\begin{equation}
j_{\ssAH}^i=e^2\epsilon^{ij} {\cal E}_j\int \frac{d^2p} {(2\pi\hbar)^2}G^{xy}f(x,p,t) \equiv \sigma_{\ssAH}\epsilon^{ij}{\cal E}_j , \nonumber 
\end{equation}
where $\sigma_{\ssAH}$ denotes the anomalous Hall conductivity.
For electrons obeying Fermi-Dirac distribution at zero temperature, 
the ground state distribution is  given by the theta-function at the Fermi energy  $E_F:$  $f=\theta(E_F-E).$ Thus,  the anamolous Hall conductivity reads
\begin{equation}
\sigma_{\ssAH}=e^2\int_{E<E_F} \frac{d^2p} {(2\pi\hbar)^2}G^{xy} . \nonumber 
\end{equation}
On the other hand the first Chern number, which is a topological invariant, is defined  by 
\begin{equation}
N_1=\frac{1}{2\pi\hbar}\int d^2p G^{xy}. \nonumber 
\end{equation} 
Therefore, one concludes that the anomalous Hall conductivity 
$$\sigma_{\ssAH}=\frac{e^2}{2\pi\hbar}N_1,$$
is a topological invariant.

\setcounter{equation}{0}

\section{Spin Chern Number vs Spin Hall Conductivity}
\label{SpinCurrent}

The semiclassical currents of the electrons obeying Dirac-like equations should  be defined in terms of the velocity which is weighted with the correct measure  $\dot{\tilde{x}}^a\tilde{w}_{1/2}.$ Recall that it is a matrix in spin indices.  We only deal  with the spin current generated by the third component of spin  $S_z,$ though any spin component can be studied similarly. The most convenient representation is 
\begin{equation}
S_z=\begin{pmatrix} I &0 \\\ 0 & -I\end{pmatrix},
\label{szre}
\end{equation}
where the dimension of the unit matrix $I$ depends on the system considered.
To define  the spin current one also needs to introduce the ground state distribution  functions $f^\uparrow (x,p,t)$ and $f^\downarrow (x,p,t)$ for the electrons with spin-up and spin-down. In the representation (\ref{szre}) we  can define the distribution  matrix  by
$$
f=\begin{pmatrix} f^{\uparrow} &0 \\\ 0 & f^\downarrow\end{pmatrix},
$$
where the   unit matrix $I$ is suppressed.  
Now, the appropriate choice for the semiclassical spin current seems to be
\begin{equation}
\label {spincurrent}
j_{z}^a=\frac{\hbar}{2}\int \frac{d^dp}{(2\pi \hbar)^d} \Tr\left[ S_z\dot{\tilde{x}}^a\tilde{w}_{1/2}f\right].
\end{equation}
Basis of the matrix representation  are the positive energy  solutions of the underlying Dirac-like equation (see Appendix A). If they are not   eigenfunctions of the spin matrix $S_z$ simultaneously, definition (\ref{spincurrent}) does not make sense.
Hence, to adopt (\ref{spincurrent}) as the definition of the spin current we  should choose the basis functions with a definite spin.
Once this is done we can set the  ground state distribution functions to unity by restricting our integrals to energies less than the Fermi energy. However, this is already the case because we deal with the wave packet composed of the positive energy solutions. 
Now, in $d=2,$ let us consider the  
 spin Hall current which results from the last term in (\ref{eom2}):
\begin{equation}
\nonumber
j_{\ssSH}^i=e\frac{\hbar}{2}\int \frac{d^2 p}{(2\pi\hbar)^2} \Tr\left[ S_zG^{ij} f\right] {\cal E}_j \equiv \sigma_{\ssSH} \epsilon^{ij} {\cal E}_j .
\end{equation} 
We are obliged to choose the basis which are spin eigenvalues so that spin Hall conductivity can be expressed as
$$
\sigma_{\ssSH}= \frac{e}{2\pi} C_s,
$$
where the spin Chern number 
\begin{equation}
\label{cs}
C_s=\frac{1}{2}(N^\uparrow -N^\downarrow),
\end{equation} 
is one half the difference of the spin-up and spin-down first Chern numbers defined by
\begin{equation}
N^{\uparrow ,\downarrow }=\frac{1}{2\pi\hbar} \int d^2p \ \Tr  G^{xy}_{\uparrow ,\downarrow }.
\label{cn1} \nonumber
\end{equation}
We demonstrated that the spin Hall conductivity is given by the spin Chern number (\ref{cs}), which  is a topological invariant characterizing the spin Hall effect. Hence,  it will be the main contribution to the spin Hall conductivity if the spin Hall phase exists.
This is the main conclusion of this work. In the following section we will illustrate this formalism by applying it to   the Kane-Mele model of graphene which is also known as  Chern insulator in $2+1$  dimensions.

\setcounter{equation}{0}
\section{Kane-Mele Model }
\label{KM}

Time reversal invariant $2+1$ dimensional topological insulator can be formulated as the spin Hall effect in graphene within the 
Kane-Mele model described by the Hamiltonian
\begin{equation}
\label{Hfull}
H= \vF \sigma_x\tau_zp_x+\vF\sigma_yp_y+\dso\sigma_z\tau_zs_z+\lambda_{\ssR}(\sigma_x\tau_zs_y-\sigma_ys_x).
\end{equation}
It is the effective theory of electrons on graphene with the Fermi velocity $\vF .$
The intrinsic  and Rashba spin-orbit coupling constants are denoted by $\dso$ and $\lambda_{\ssR},$ respectively.
 $\sigma_{x,y,z}$   are the Pauli matrices  in the representation $\sigma_z=\diag (1,-1),$
which act on the states of sublattices.
$\tau_z=\diag (1,-1),$ labels the states at the Dirac points (valleys) $K$ and $K',$ and   
the Pauli matrices,  $s_{x,y,z}$ act on the real spin space in the representation where the third component is diagonal
$s_z=\diag (1,-1).$ 

 The main difference between the Kane-Mele model with and without the Rashba spin-orbit coupling term lies in whether the third component of spin is a good quantum number or not.
In the former case $s_z$ is conserved and application of the semiclassical approach is straightforward.
However, also in the latter case the spin Hall conductivity is non-vanishing with the condition $\dso > \lambda_{\ssR} . $
We will illustrate how the semiclassical formulation can be applied in both cases and demonstrate that 
main contribution to the spin Hall conductivity is always given by  the spin Chern number defined in \cite {prodan}.

\subsection{ The $\lambda_{\ssR}=0$ Case}

In this case the Hamiltonian is 
\begin{equation}
\label{HKM}
H^{\ssSO}= \vF \sigma_x\tau_zp_x+\vF\sigma_yp_y+\dso\sigma_z\tau_zs_z.
\end{equation}
The 
Foldy-Wouthuysen transformation \cite{dayiyunt} 
$$U=\frac{\sigma_z\tau_zs_zH^{\ssSO}+E}{\sqrt{2E(E+\dso)}},$$
 diagonalizes the Hamiltonian (\ref{HKM}):
\begin{equation}
\nonumber
 {\cal H}^{\ssSO}=UH^{\ssSO}U^{\dagger}=E\sigma_z\tau_zs_z,
\end{equation}
where $E=\sqrt{\vF^2p^2+\dso^2}.$ $U$ can be employed to acquire the eigenfunctions of (\ref{HKM}) as
$$
u^{(\alpha)}(p) =U^\dagger v^{(\alpha)},
$$
where $v^{(1)}=(1,0\cdots,0)^T,\cdots, v^{(8)}=(0,0\cdots,8)^T.$
The Hamiltonian projected on positive energy eigenstates in the presence of the external electric field  $\bm {{\cal E}}$ is
$$H_0^{\ssSO} = (E+e \bm{{\cal E}}\cdot  \bm x){\bm 1}_{\tau}{\bm 1}_{s} .$$
In the rest of this section we will keep  the unit matrices explicit.
The Berry gauge field can be shown to be 
$$ A^i= \frac {\hbar\vF^2}{2E(E+\dso)}\epsilon^{ij}p_j{\bm 1}_{\tau}s_z. $$
Hence, the corresponding Berry curvature, $G^{xy}=(\partial_{p_x}A^y-\partial_{p_y}A^x)$, is
$$G^{xy}=-\frac {\hbar\vF^2\dso}{2E^3}{\bm 1}_{\tau}s_z.$$
In the absence of a magnetic field the phase space measure (\ref{eom1}) is trivial: $\tilde{w}_{1/2} =1.$ Thus, 
 the equations of motion (\ref{eom2})-(\ref{eom3}) yield 
\begin{eqnarray}
\dot{\tilde{x}}^i &=& -\frac{\vF^2g^{ij}p_j}{E}\tau_zs_z+e\epsilon^{ij}{\cal E}_jG^{xy} , \nonumber\\
\dot{\tilde{p}}_i &=& e{\cal E}_i{\bm 1}_{\tau}{\bm 1}_{s}  , \nonumber
\end{eqnarray}
where $g^{ij}$ is the Euclidean metric. In the representation which we adopted the third component of spin becomes
\begin{equation}
\label{spin}
S_z={\bm 1}_{\tau} s_z .
\end{equation}
Note that   $u^{(\alpha)}$ are also the  eigenstates of the spin matrix (\ref{spin}).
Therefore, the distribution matrix  $f={\bm 1}_{\tau}\diag (f^\uparrow  ,f^\downarrow  )$ is adequate
  to define the spin current by
\begin{equation}
j_{z}^i=\frac{\hbar}{2}\int \frac{d^2p} {(2\pi\hbar)^2}\Tr\left[ S_z\sqrt{w}\dot{\tilde{x}}^i
f\right] . \nonumber
\end{equation}
It yields the spin Hall current
$j_{\ssSH}^{i}=\sigma_{\ssSH}\epsilon^{ij}{\cal E}_j$, where  the spin Hall conductivity is given by 
\begin{equation}
\label{sigmashKM}
\sigma_{\ssSH}=\frac{e\hbar}{2} \int \frac{d^2p} {(2\pi\hbar)^2}\Tr\left[S_z G^{xy}\right].
\end{equation}
Let us decompose (\ref{sigmashKM}) such that the contributions arising from  spin subspace and $K$, $K'$ valleys become apparent.
One can easily observe that 
\begin{eqnarray}
\sigma_{\ssSH} &=& \frac{e\hbar}{2}\int \frac{d^2p}{(2\pi\hbar)^2}( G^{xy}_{\uparrow K}- G_{\downarrow K}^{xy}+ G_{\uparrow K'}^{xy}- G_{\downarrow K'}^{xy}) \nonumber \\
&=&\frac{e}{4\pi}( N_1^{\uparrow K}-N_1^{\downarrow K} + N_1^{\uparrow K'}-N_1^{\downarrow K'}). \nonumber
\end{eqnarray}
Each  contribution is  associated with the  first Chern number of the related subspace. This has been observed in \cite{dayiyunt} where the related Chern numbers were calculated.
We conclude that the spin Hall conductivity is proportional to the sum of the spin Chern number of the $K$ valley, $C^{K}_s$ and the spin Chern number of the $K'$ valley, $C^{K'}_s:$
$$
\sigma_{\ssSH}=\frac{e}{2\pi}(C^{K}_s+C^{K'}_s) =\frac{e}{2\pi}C_s
                        =\frac{e}{2\pi}
                        .
$$
In the absence of  Rashba term 
we  defined the spin current   	straightforwardly since the Hamiltonian (\ref{HKM}) commutes with $s_z.$ 

\subsection{ The $\lambda_{\ssR}\neq0$ Case}
\label{kmwr}

\noindent
 Although, in the presence of the Rashba term  $s_z$ does not commute with the Hamiltonian  (\ref{Hfull}),  the spin Hall effect still exists for 
$\dso > \lambda_{\ssR} $ \cite{km,ssth,effRas}. However, the semiclassical calculation is not clear as we discussed in Section \ref{SpinCurrent}. There we also discussed the correct definition of spin current. Nevertheless,  before proceeding as indicated in Section \ref{SpinCurrent}  let us carry on with the computation of the Berry gauge field naively using the positive energy eigenfunctions of (\ref{Hfull}).

The $K$ and $K'$ subspaces corresponding to $\tau_z=\pm 1$ yield the same energy eigenvalues and eigenstates which are  presented in Appendix B. Thus, it is sufficient to consider  only the $4 \times 4$ Hamiltonian in $K$ subspace denoted by $H_{K}:$ 
$$ H_{\ssK}\Phi_{\alpha}=E_{\alpha}\Phi_{\alpha},$$
where $\alpha=1,..,4$ and the energy eigenvalues $E_{\alpha}$ are
\begin{eqnarray}
E_1&=& \lambda+ \sqrt{(\dso-\lambda)^2+\vF^2p^2}, E_2 =-\lambda+ \sqrt{(\dso+\lambda)^2+\vF^2p^2} ,\nonumber\\
E_3 &=&\lambda- \sqrt{(\dso-\lambda)^2+\vF^2p^2} ,E_4 =-\lambda- \sqrt{(\dso+\lambda)^2+\vF^2p^2}. \label{energy}
\end{eqnarray}
We deal with the coupling constants satisfing $\dso>2\lambda_{\ssR}$, so that $E_1,E_2$ and $E_3,E_4$ are  positive and  negative, respectively. 

The diagonalized Hamiltonian  is ${\cal H}_{\ssK}^{\Phi} =\diag (E_1,E_2,E_3,E_4 ).$
When we project on the positive energy eigenstates and take both of the contributions coming from the $K$ and $K'$ subspaces, the Hamiltonian becomes
$$
H^{\Phi}_0={\bm 1}_{\tau}\begin{pmatrix}E_1&0 \\\ 0& E_2\end{pmatrix} + e \bm {{\cal E}} \cdot  \bm x{\bm 1}_{\tau} {\bm 1}_{s}.
$$
The Berry gauge field  turns out to be Abelian:
$$
A^i_{\Phi}=\hbar\epsilon^{ij}\frac{p_j}{p^2}{\bm 1}_{\tau}\begin{pmatrix} -1 & 2N_1N_2 \\\ 2N_1N_2 & -1\end{pmatrix} .
$$
The corresponding Berry curvature $G_{\Phi}^{xy}=(\partial_{p_x}A_{\Phi}^y-\partial_{p_y}A_{\Phi}^x),$ can easily be computed  as
$$G_{\Phi}^{xy}={\bm 1}_{\tau}\begin{pmatrix}0&-\frac{2\hbar}{p}\partial_p(N_1N_2)\\\ -\frac{2\hbar}{p}\partial_p(N_1N_2)&0 \end{pmatrix}.$$
According to (\ref{eom1})-(\ref {eom3}) the equations of motion calculated in the energy eigenfunction basis  are
\begin{eqnarray}
\dot{\tilde{x}}^i &=&- {\bm 1}_{\tau}\vF^2g^{ij}p_j\begin{pmatrix}\frac{1}{E_1-\lambda}&0\\\ 0&\frac{1}{E_2+\lambda}\end{pmatrix}+2\frac{N_1N_2}{p^2}\epsilon^{ij}p_j(E_1-E_2){\bm 1}_{\tau}s_{y}+e\epsilon^{ij}{\cal E}_j{\bm 1}_{\tau}G_{\Phi}^{xy} ,\nonumber \\
\dot{\tilde{p}}_i &=& e{\cal E}_i{\bm 1}_{\tau}{\bm 1}_{s} , \nonumber
\end{eqnarray}
where we set $\tilde{w}_{1/2} =1.$ 

The spin current cannot be defined by (\ref{spincurrent}) with a diagonal $f.$ Choosing it diagonal would lead to a vanishing spin Hall current due to the fact that
$$
\Tr\left[S_z G_{\Phi}^{xy}\right] =0.
$$
The difficulty stems from the fact that energy eigenfunctions are not simultaneously eigenstates of the spin operator  $S_z=\diag ( s_z,s_z ) . $
In $K$ subspace  eigenstates  of the spin operator $S_z,$ constructed from the energy eigenstates $\Phi_\alpha,$ are 
\begin{eqnarray}
\Psi_1 &=& \frac{1}{\sqrt 2}(\Phi_{1} +\Phi_{2}), \Psi_2 = \frac{1}{\sqrt 2}(\Phi_{1} -\Phi_{2}),\nonumber \\
\Psi_3 &=& \frac{1}{\sqrt 2}(\Phi_{3} +\Phi_{4}), \Psi_4 = \frac{1}{\sqrt 2}(\Phi_{3} -\Phi_{4}).\nonumber
\end{eqnarray}
 $\Psi_1, \Psi_2$ and $\Psi_3, \Psi_4$  correspond to positive and negative energy sectors, respectively. They satisfy 
$$ S_z\Psi_{1,3} =  \Psi_{1,3} ,\ S_z\Psi_{2,4} =-  \Psi_{2,4}. $$
The Hamiltonian in $\Psi_{\alpha}$ basis is obtained by the transformation
$${\cal H}^{\Psi}_{\ssK}=U_{\Psi}H_{\ssK}U^{\dagger}_{\Psi}, $$
 where $U^{\dagger}_{\Psi}=\begin{pmatrix}\Psi_1&\Psi_2&\Psi_3&\Psi_4 \end{pmatrix}$.
Notice that $U_{\Psi}$ is related to the unitary transformation that diagonalizes $H_{\ssK}$, denoted by $U_{\Phi}$, via $U_{\Psi}=RU_{\Phi}$, where $R= \begin{pmatrix}\tilde{R}&0 \\ 0&\tilde{R}\end{pmatrix}$ with
$$ \tilde{R}= \frac{1}{\sqrt 2}\begin{pmatrix} 1&1 \\ 1&-1\end{pmatrix}.
$$
Thus we acquired
\begin{eqnarray}
{\cal H}^{\Psi} &=& ({\bm 1}_{\tau}R){\cal H}^{\Phi}(R{\bm 1}_{\tau})       \nonumber \\
                    &=& \frac{1}{2}\begin{pmatrix} E_1+E_2&E_1-E_2&0&0 \\ E_1-E_2&E_1+E_2&0&0\\0&0&E_3+E_4&E_3-E_4\\0&0&E_3-E_4&E_3+E_4\end{pmatrix}{\bm 1}_{\tau}  .\nonumber     
\end{eqnarray}
The Hamiltonian in $\Psi_{\alpha}$ basis  projected on positive energy eigenstates in the presence of external electric field $\bm{{\cal E}}$,  is 
$$H_0^{\Psi} = (E_1+E_2){\bm 1}_{\tau}{\bm 1}_s+(E_1-E_2){\bm 1}_{\tau}s_x+e \bm{{\cal E}}\cdot  \bm x{\bm 1}_{\tau} {\bm 1}_{s} .$$
The basis transformation $\tilde R $ sustains the connection between ${\bm A}_{\Psi}$ and ${\bm A}_{\Phi}$ via the relation ${\bm A}_{\Psi}=({\bm 1}_{\tau}\tilde R){\bm A}_{\Phi}(\tilde R{\bm 1}_{\tau}) ,$   so that 
$$
A^i_{\Psi} =\hbar\epsilon^{ij} \frac{p_j}{p^2}{\bm 1}_{\tau}\begin{pmatrix} -1+2N_1N_2 &0\\\ 0&-1+2N_1N_2\end{pmatrix} \equiv {\bm 1}_{\tau}\begin{pmatrix}A^i_{\uparrow}&0 \\\ 0& A^i_{\downarrow} \end{pmatrix}. $$
The corresponding Berry curvature $G^{\Psi}_{xy} = ({\bm 1}_{\tau}\tilde R) G^{\Phi}_{xy}(\tilde R{\bm 1}_{\tau}) $ is calculated as
$$G_{\Psi}^{xy}=- \frac{2\hbar}{p} \frac{\partial(N_1N_2)}{\partial p}{\bm 1}_{\tau}s_z \equiv {\bm 1}_{\tau}\begin{pmatrix} G^{xy}_{\uparrow}&0 \\\ 0&G^{xy}_{\downarrow} \end{pmatrix} . $$
Hence, the equations of motion are 
\begin{eqnarray}
\dot{\tilde{x}}^i &=& -{\bm 1}_{\tau}\vF^2g^{ij}p_j\begin{pmatrix}\frac{1}{E_1 -\lambda}+\frac{1}{E_2 +\lambda}&\frac{1}{E_1 -\lambda}-\frac{1}{E_2 +\lambda} \\\ \frac{1}{E_1 -\lambda}-\frac{1}{E_2 +\lambda}&\frac{1}{E_1 -\lambda}+\frac{1}{E_2 +\lambda}\end{pmatrix}-2\frac{N_1N_2}{p^2}\epsilon^{ij}p_j(E_1-E_2){\bm 1}_{\tau}s_{y}+e\epsilon^{ij}{\cal E}_jG_{\Psi}^{xy} ,\nonumber\\
\dot{\tilde{p}}_i &=& e{\cal E}_i {\bm 1}_{\tau}{\bm 1}_s.  \nonumber
\end{eqnarray}
The spin Hall current can now be written as
$$
j_{\ssSH}^i=\frac{e\hbar}{2}\epsilon^{ij}{\cal E}_j\int \frac{d^2p} {(2\pi\hbar)^2}\Tr\left[S_z  G_{\Psi}^{xy}f\right],
$$
where
$
f= \bm 1_\tau\diag (f^{\uparrow},f^{\downarrow}).
$
Therefore, $f$ restricts the integral to energies less than the Fermi energy and
the spin Hall conductivity becomes
$$
\sigma_{\ssSH}=\frac{e\hbar}{2} \int\frac{d^2p}{(2\pi\hbar)^2}\Big\{(G^{xy}_{K\uparrow}-G^{xy}_{K\downarrow})+(G^{xy}_{K'\uparrow}-G^{xy}_{K'\downarrow})\Big\} =\frac{e}{2\pi}(C^{K}_s+C^{K'}_s)=\frac{e}{2\pi}C_s.
$$
In \cite{yang} this spin Chern number is calculated as  $C_s=1.$ Therefore, we conclude that
$$\sigma_{\ssSH}= \frac{e}{2\pi}.$$
Indeed, in \cite{km} it was argued that the  value of the spin Hall conductivity sligtly differs from this value which  is confirmed either in terms of numerical methods \cite{ssth} or deriving the related effective theory \cite{effRas}.

\section{Discussion}
In $2+1$ dimensions we established  the anomalous Hall conductivity as well as the  spin Hall conductivity from the term linear in the electric field and the Berry curvature in $\dot{\tilde x}^i\tilde{w}_{1/2}.$ This anomalous velocity term survives in any $d+1$ spacetime dimension: 

Independent of the spacetime  dimension and the origin of the Berry curvature in the time evolution of the coordinates there is always a term which is linear in both electric field and the Berry field strength,
$$ \frac{\partial}{\partial {\cal E}_a}(\tilde{w}_{1/2}\dot{\tilde x}^b)|_{{\cal \bm E}=0}\propto G^{ab}.$$

In the basis where a certain component of spin is diagonal this term will be diagonal.  
Therefore,
procedure of calculating the spin Hall conductivity can be generalized to any dimension. However, topological origin of this conductivity should be discussed within the underlying physical system.

\setcounter{equation}{0}
\setcounter{section}{1} \null

\renewcommand{\theequation}{\Alph{section}.\arabic{equation}}
\renewcommand{\thesection}{\Alph{section}}

\section*{Appendix A}

Dirac equation possesses negative and positive energy solutions. Obviously one can form a wave packet by superposing only  positive  energy solutions. However, relativistic invariance of the Dirac theory demands to superpose both positive and negative solutions. Nevertheless by ignoring the relativistic momenta one can only deal with a wave packet composed of positive energy solutions. Indeed this is the starting point of the semiclassical approximation.  We denote the spinor corresponding to a positive energy solution of Dirac equation by $u^{(\alpha)}({\bm p,\bm x_c})$,  
which is a function of the momentum ${\bm p}$, and the position of the wave packet center in coordinate space ${\bm x_c}:$
$$H_0(\bm p )u^{(\alpha)}(\bm p,\bm x_c)=E_\alpha u^{(\alpha)}(\bm p,\bm x_c);\ E_\alpha > 0.$$
 The normalization is  $ u^{\dagger(\alpha)}({\bm p,x_c})u^{(\beta)}({\bm p,x_c})=\delta_{\alpha\beta}$. Let us consider the following wave packet obtained by superposing only positive energy solutions labeled by the superscript $\alpha$,
\begin{equation}
\label{wavepacket}
\Psi_{\bm{x}}\equiv \Psi_{\bm{x}} (\bm{p}_c, \bm{x}_c)=\int [dp]|a({\bm p},t)|e^{-i\gamma({\bm p},t)}\sum_{\alpha}\xi_{\alpha}\psi_{\bm{x}}^{(\alpha)}({\bm p,\bm{x}_c}) ,
\end{equation}
where $[dp]$  denotes the measure of the $d$ dimensional momentum space. The distribution $|a({\bm p},t)|e^{-i\gamma({\bm p},t)}$ has a peak at  the wave packet center $\bm{p}_c$ and satisfies  $\int |a|^2[dp]=1.$ The expansion coefficients
 $\xi_{\alpha}$ are also  normalized, $\sum_{\alpha}| \xi_{\alpha}|^2=1$. $\psi_{\bm{x}}^{(\alpha)}({\bm p,\bm x_c}) $  is composed of two parts
\begin{equation}
\label{psimalpha}
\psi_{\bm{x}}^{(\alpha)}({\bm p,\bm x_c})=u^{(\alpha)}({\bm p,\bm x_c})\phi_{\bm{x}}(\bm{p}),
\end{equation}
with
$$\phi_{\bm{x}}(\bm{p}) =\frac{1}{(2\pi)^{d/2}}e^{-i\bm p \cdot \bm x}.$$
The normalization is
$$
\int [dp] \phi^\star_{\bm{x}}(\bm{p})\phi_{\bm{y}}(\bm{p}) =\delta(\bm{x}-\bm{y}).
$$
When the position operator,
 $\hat {\bm{x}}$ acts on $\phi_{\bm{x}}(\bm{p})$ we get
$$
\hat{{\bm{x}}}\phi_{\bm{x}}(\bm{p})=i\frac{\partial}{\partial \bm{p}}\phi_{\bm{x}}(\bm{p})={\bm{x}}\phi_{\bm{x}}(\bm{p}).
$$
The completeness relation is $\int [dx] \phi^\star_{\bm{x}}(\bm{p})\phi_{\bm{x}}(\bm{q})=\delta(\bm{p}-\bm{q})$. As a result of these definitions, (\ref{psimalpha}) has the following normalization 
\begin{equation}
\label{normalization}
\int [dx]\psi_{\bm{x}}^{\dagger(\alpha)}({\bm p,\bm x_c})\psi_{\bm{x}}^{(\beta)}({\bm q,\bm x_c})=\delta_{\alpha\beta}\delta(\bm p-\bm q).
\end {equation}
We would like to calculate the expectation value of the position operator over the wave packet (\ref{wavepacket}), which is equivalent to ${\bm x_c}=\int [dx]\Psi^\dagger_{\bm{x}}\bm{\hat{x}}\Psi_{\bm{x}} $.
The calculation proceeds as follows; we first calculate $\bm{\hat{x}}\Psi_{\bm{x}}$, in which we use 
$$
\hat{{\bm{x}}}\psi_{\bm{x}}^{(\alpha)}=u^{(\alpha)}({\bm p,\bm x_c})\hat{{\bm{x}}}\phi_{\bm{x}}(\bm{p})=u^{(\alpha)}({\bm p,\bm x_c}){\bm{x}}
\phi_{\bm{x}}(\bm{p})=u^{(\alpha)}({\bm p,\bm x_c})(i\frac{\partial}{\partial \bm{p}})\phi_{\bm{x}}(\bm{p}).
$$
Now, integrating by parts we obtain
\begin{eqnarray}
\hat{{\bm{x}}}\Psi_{\bm{x}} &=&-i\int [dp] \frac{\partial |a(\bm p,t)|}{\partial \bm p}e^{-i\gamma({\bm p},t)}\sum_{\alpha}\xi_{\alpha}u^{(\alpha)}({\bm p,\bm x_c})\phi_{\bm{x}}(\bm{p}) \nonumber\\ 
&-&\int [dp]|a(\bm p,t)|\frac{\partial \gamma({\bm p},t)}{\partial \bm p} e^{-i\gamma({\bm p},t)}\sum_{\alpha}\xi_{\alpha}u^{(\alpha)}({\bm p,\bm x_c})\phi_{\bm{x}}(\bm{p}) \nonumber \\
&-& i\int [dp]|a(\bm p,t)|e^{-i\gamma({\bm p},t)}\sum_{\alpha}\xi_{\alpha}\frac{\partial u^{(\alpha)}(\bm p, \bm x_c)}{\partial \bm p}\phi_{\bm{x}}(\bm{p}). \nonumber
\end{eqnarray}
Then we reach the following result
\begin{eqnarray}
\int [dx]\Psi^\dagger_{\bm{x}}\hat{x}\Psi_{\bm{x}} &=&- i \int [dp]|a(\bm p,t)|\frac{\partial |a(\bm p,t)|}{\partial \bm p} \label{secondstep} \\
&-& \int [dp]|a(\bm p,t)|^2\frac{\partial \gamma({\bm p},t)}{\partial \bm p}
-i\int [dp]|a(\bm p,t)|^2\sum_{\alpha,\beta}\xi^*_{\beta}u^{\dagger(\beta)}({\bm p,\bm x_c})\frac{\partial u^{(\alpha)}({\bm p,\bm x_c})}{\partial \bm p}\xi_{\alpha} .  \nonumber
\end{eqnarray}
The first term vanishes since $\int |a|^2[dp]=1$. The second and the third terms are obtained using (\ref{normalization}). The distribution has the mean momentum, $\bm p_c$ defined through the integral
$$ \bm p_c=\int [dp]\bm p|a(\bm p,t)|^2.$$
Moreover, for any function $f(\bm p),$ we get
\begin{equation}
\label{fqc}
f(\bm p_c)= \int [dp]f(\bm p) |a(\bm p,t)|^2.
\end{equation} 
Using the definition (\ref{fqc}) in (\ref{secondstep}), we obtain
\begin{equation}
\label{xc}
\bm {x_c}=- \frac{\partial\gamma_c}{\partial {\bm p_c}} -i\sum_{\alpha,\beta}\xi^*_{\beta} u^{\dagger(\beta)}(\bm p_c,\bm x_c)\frac{\partial }{\partial {\bm p_c}}u^{(\alpha)}(\bm p_c,\bm x_c) \xi_{\alpha},
\end{equation}
where $\gamma_c\equiv\gamma(\bm p_c,t)$.
We would like to obtain the one-form $\eta,$ which is defined through $d{\cal S}:$
$$
d{\cal S}\equiv\int [dx]\Psi^\dagger_{\bm{x}}\left( -id-H_0dt\right) \Psi_{\bm{x}}=- d\gamma_c +\sum_{\alpha\beta}\xi^*_{\alpha} \eta^{\scriptscriptstyle{\alpha\beta}}\xi_{\beta}.
$$
We start by computing 
\begin{eqnarray}
d\Psi_{\bm{x}}&=&dt \frac{\partial\Psi_{\bm{x}}}{\partial t}+d{\bm x}_c\frac{\partial\Psi_{\bm{x}}}{\partial \bm{x}_c} \nonumber\\
&=&dt\int [dp]\frac{\partial}{\partial t}|a(\bm p,t)|e^{-i\gamma(\bm p,t)}\sum_{\alpha}\xi_{\alpha}u^{(\alpha)}(\bm p,\bm x_c)\phi_{\bm{x}}(\bm{p}) \nonumber\\&-&idt\int [dp]|a(\bm p,t)|\frac{\partial \gamma(\bm p,t)}{\partial t}e^{-i\gamma(\bm p,t)}\sum_{\alpha}\xi_{\alpha}u^{(\alpha)}(\bm p,\bm x_c)\phi_{\bm{x}}(\bm{p})
\nonumber\\&+&d{\bm x}_c\int [dp]|a(\bm p,t)|e^{-i\gamma(\bm p,t)}\sum_{\alpha}\xi_{\alpha}\frac{\partial }{\partial \bm x_c}u^{(\alpha)}(\bm p,\bm x_c)\phi_{\bm{x}}(\bm{p}) .\nonumber
\end{eqnarray}
So that we obtain 
$$
\int [dx]\Psi^\dagger_{\bm{x}}id\Psi_{\bm{x}} = dt\frac{\partial \gamma_c}{\partial t} + id{\bm x}_c\sum_{\alpha\beta}\xi^*_{\beta}u^{\dagger(\beta)}(\bm p_c,\bm x_c)\frac{\partial}{\partial \bm x_c} u^{(\alpha)}(\bm p_c,\bm x_c)\xi_{\alpha} .
$$
To transform the first term, we use  $d\gamma_c=dt\frac{\partial \gamma_c}{\partial t}+d{\bm p}_c\frac{\partial \gamma_c}{\partial \bm p_c}$ and (\ref{xc}):
\begin{equation}
 \int [dx]\Psi^\dagger_{\bm{x}}id\Psi_{\bm{x}} =  d\gamma_c +d{\bm{p}}_c\cdot\bm{x}_c+id{\bm x}_c\sum_{\alpha\beta}\xi^*_{\beta}u^{\dagger(\beta)}\frac{\partial }{\partial \bm x_c}u^{(\alpha)}\xi_{\alpha} +id{\bm p}_c\sum_{\alpha\beta}\xi^*_{\beta}u^{\dagger(\beta)}\frac{\partial }{\partial \bm p_c}u^{(\alpha)}\xi_{\alpha}  \nonumber
\end{equation}
Here is a convenient point to define the following matrix valued Berry gauge fields
\begin{eqnarray}
 iu^{\dagger(\alpha)}(\bm p_c,\bm x_c)\frac{\partial }{\partial {\bm x_c}} u^{(\beta)}(\bm p_c,\bm x_c)&=&\bm a^{\alpha\beta}, \nonumber\\
 iu^{\dagger(\alpha)}(\bm p_c,\bm x_c)\frac{\partial }{\partial {\bm p_c}} u^{(\beta)}(\bm p_c,\bm x_c)&=&\bm A^{\alpha\beta} . \nonumber
\end{eqnarray}
The Dirac-like free Hamiltonian only depends on the derivatives with respect to $\bm x ,$ so that we get
$$
\int [dx]\Psi^\dagger_{\bm{x}}H_0(\frac{\partial}{\partial \bm{x}} ) \Psi_{\bm{x}}= \sum_{\alpha\beta}\xi^*_{\alpha} E_\alpha (\bm{p}_c) \delta^{\alpha\beta}\xi_{\beta}.
$$
Thus, by defining $H_0^{\alpha\beta}=E_{\alpha}\delta^{\alpha\beta},$ we obtain 
$$
d{\cal S}=\int [dx]\Psi^\dagger_{\bm{x}}\left(- id-H_0dt\right)\Psi_{\bm{x}}=-d\gamma_c -\sum_{\alpha\beta}\xi^*_{\alpha}\Big(d{\bm{p}}_c\cdot\bm{x}_c\delta^{\alpha\beta}+d{\bm x}_c\bm a^{\alpha\beta}+d{\bm p}_c\bm A^{\alpha\beta} + H_0^{\alpha\beta}dt\Big)\xi_{\beta} .
$$
Therefore, we can define the matrix valued one-form $\eta^{\alpha\beta}$ by,
$$
\eta^{\alpha\beta}= -\delta^{\alpha\beta}\bm{x}_c\cdot d\bm{p}_c-\bm a^{\alpha\beta}\cdot d{\bm x}_c-\bm A^{\alpha\beta}\cdot d{\bm p}_c - 
 H_0^{\alpha\beta} dt.
$$
It governs the dynamics of the wave-packet. Note that under the unitary transformation of the basis $\tilde{u}^{(\alpha)}=U_{\beta\alpha}u^{(\beta)},$ the  one-form $\eta$ transforms as $\tilde \eta=U\eta U^{\dagger}.$
\section*{Appendix B}
 The Hamiltonian for the $K$ subspace is obtained from (\ref{Hfull}) by setting $\tau_z=1:$
 $$H_{\ssK}=\begin{pmatrix} \dso s_z& {\bm 1}_{s}\vF(p_x-ip_y)+\lambda_{\ssR}(s_y+is_x) \\\ {\bm 1}_{s}\vF(p_x+ip_y)+\lambda_{\ssR}(s_y-is_x) &-\dso s_z \end{pmatrix}. $$
The eigenstates of $H_{\ssK}$ corresponding to the energy eigenvalues (\ref{energy}) can be shown to be
$$\Phi_1=N_1\begin{pmatrix} i\frac{p_x-ip_y}{p_x+ip_y}\\\ \frac{E_1-\dso}{\vF(p_x+ip_y) }\\ -i\frac{E_1-\dso}{\vF(p_x+ip_y) }\\ 1\end{pmatrix},\Phi_2=N_2\begin{pmatrix} -i\frac{p_x-ip_y}{p_x+ip_y}\\\ \frac{E_2-\dso}{\vF(p_x+ip_y) }\\ -i\frac{E_2-\dso}{\vF(p_x+ip_y) }\\ 1\end{pmatrix},
\Phi_3=N_3\begin{pmatrix} i\frac{p_x-ip_y}{p_x+ip_y}\\\ \frac{(E_3-\dso)}{\vF(p_x+ip_y) }\\ -i\frac{(E_3-\dso)}{\vF(p_x+ip_y) }\\ 1\end{pmatrix},\Phi_4=N_4\begin{pmatrix}- i\frac{p_x-ip_y}{p_x+ip_y}\\\ \frac{(E_4-\dso)}{\vF(p_x+ip_y) }\\ i\frac{(E_4-\dso)}{\vF(p_x+ip_y) }\\ 1\end{pmatrix},$$
where the normalizations are $N_{\alpha}(p)=\frac{\vF p}{\sqrt{2(\vF^2p^2+(E_{\alpha}-\dso)^2)}}$.

 When $\tau_z=-1$ is taken in (\ref{Hfull}), the Hamiltonian for the $K'$ valley  is obtained:
$$H_{\ssKprime}=\begin{pmatrix} -\dso s_z& -{\bm 1}_{s}\vF(p_x+ip_y)-\lambda_{\ssR}(s_y-is_x) \\\ -{\bm 1}_{s}\vF(p_x-ip_y)-\lambda_{\ssR}(s_y+is_x) &\dso s_z \end{pmatrix}.$$
 The  eigenstates of $H_{\ssKprime}$ are as follows,
$$\Phi_5=N_1\begin{pmatrix}- i\frac{E_1-\dso}{\vF(p_x+ip_y)}\\\ 1\\ i\frac{p_x-ip_y}{p_x+ip_y }\\ -\frac{E_1-\dso}{\vF(p_x+ip_y)}\end{pmatrix},\Phi_6=N_2\begin{pmatrix} i\frac{E_2-\dso}{\vF(p_x+ip_y)}\\\ 1\\- i\frac{p_x-ip_y}{p_x+ip_y }\\ -\frac{E_2-\dso}{\vF(p_x+ip_y)}\end{pmatrix},
\Phi_7=N_3\begin{pmatrix}- i\frac{E_3-\dso}{\vF(p_x+ip_y)}\\\ 1 \\ i\frac{p_x-ip_y}{p_x+ip_y}\\ -\frac{E_3-\dso}{\vF(p_x+ip_y)}\end{pmatrix},\Phi_8=N_4\begin{pmatrix} i\frac{E_4-\dso}{\vF(p_x+ip_y)}\\\ 1 \\ -i\frac{p_x-ip_y}{p_x+ip_y}\\ -\frac{E_4-\dso}{\vF(p_x+ip_y)} \end{pmatrix}.$$
The corresponding energy eigenvalues are given by (\ref{energy}) since  $E_5=E_1$, $E_6=E_2$, $E_7=E_3$, $E_8=E_4$.

\newpage

\newcommand{\PRL}{Phys. Rev. Lett. }
\newcommand{\PRB}{Phys. Rev. B }
\newcommand{\PRD}{Phys. Rev. D }

\end{document}